\documentclass[%
reprint,
 amsmath,amssymb,
 aps,
pra,
]{revtex4-2}
\usepackage{xcolor}
\usepackage[T1]{fontenc}
\usepackage{graphicx}
\usepackage{dcolumn}
\usepackage{bm}

\begin{document}


\title{Induced Berry Connection and Photonic Spin Hall Effect in Optical Dirac Theory}

\author{Lili Yang$^{1}$}
\author{Longlong Feng$^{1}$}
\email{flonglong@mail.sysu.edu.cn}
\affiliation{$^{1}$ School of Physics and Astronomy, Sun Yat-Sen University, Zhuhai 519082, China}

\date{\today}

\begin{abstract}

Within the framework of optical Dirac theory, we present a field-theoretical model of spin-orbit interaction and photonic spin/orbit Hall effects. Our approach reformulates light propagation along helical paths as solving the Maxwell equations in a ray-based curvilinear coordinate system. This system can alternatively be interpreted as a spin-degenerate medium with antisymmetric elements in the dielectric tensor, corresponding to the spin-1 excitation mode characterized by a rotational dipole vector. We show that, at leading order, the resulting effective Hamiltonian is equivalent to the Maxwell theory in a uniformly rotating frame, incorporating both spin-rotation and orbit-rotation coupling. This rotation arises from the torsion of helical paths, manifesting as extrinsic orbital angular momentum (EOAM) of photons. Notably, the spin angular momentum (SAM) of the photon and its intrinsic orbital angular momentum (IOAM) contribute jointly to the geometric phase. In the Heisenberg picture, the spin and orbital Hall effects naturally emerge from interaction terms in the ray equations. Furthermore, we find that the transverse spin of evanescent waves couples with EOAM, revealing that the geometric phase of elliptically polarized light differs from that of circularly polarized light. This distinction underscores the role of spin-orbit coupling in modifying phase accumulation.

\end{abstract}

\maketitle

\section{Introduction}

Polarization refers to the orientation of an electromagnetic wave's electric/magnetic field vector. In the context of quantum theory, it corresponds to the spin angular momentum of photons, an intrinsic degree of freedom of light encoded in the classical Maxwell theory. Adapting our understanding of electromagnetic waves to incorporate the concept of angular momentum is critical for comprehending the intricate dynamics at play. Besides the spin angular momentum (SAM) of an electromagnetic wave packet, we have appreciated another two pivotal components: intrinsic orbital angular momentum(IOAM) associated with optical vortices, which characterizes the inherent rotation of light around its axis, and extrinsic orbital angular momentum (EOAM) corresponds to the helical optical trajectory, extending beyond its inherent properties to encompass its path through space \cite{10.1117/12.317704,Allen1992OrbitalAngularMomentum,Bekshaev2008ParaxialLB,Bekshaev_2011,PhysRevLett.88.053601,PhysRevLett.90.133901,Bliokh2015TransverseAL}. Interactions among these three angular momenta give rise to a diverse range of optical phenomena and can be manipulated to control various aspects of optical systems. 

The spin-orbit interaction (SOI) of light manifests itself through the interplay and mutual conversion between three types of angular momentum\cite{Onoda2004HallEO,Bliokh2008GeometrodynamicsOS,Fedoseyev2008TransformationOT,Bliokh_2009,PhysRevLett.103.100401,articleAiello,PhysRevA.82.063825,Bliokh_2014,Marrucci_2011,Aiello2015FromTA,Bliokh2021SpatiotemporalVP}, a family of fundamental phenomena that arises directly from the solutions of Maxwell's equations, particularly when dealing with wavelength-scale micro- and nanostructures. Obviously, a full treatment of the SOI phenomenon on these scales requires vector-wave mechanics. This motivation has led us to develop an extended Dirac theory for optical fields in generic media, which was found to be a non-Hermitian chiral extension of massive fermions with anomalous magnetic moments moving in an external pseudo-magnetic field \cite{Feng2022FourVectorOD}. In this approach, an optical medium prescribed by a symmetric dielectric tensor can be decomposed in terms of the monopole, dipole, and quadrupole components, corresponding to the spin-0 scalar mode, spin-1 vector mode, and spin-2 tensor mode, respectively. While coupling to optical fields, those virtual excitation modes ensure the conservation of total angular momentum (TAM). On the other hand, the chiral $\gamma_5$ extension leads to a scenario of an anomalous magnetic momentum in an equivalent magnetic field, a synthetic one in a non-spin-degenerate medium. 

This study aims to develop a field-theoretical representation of spin-orbit interaction (SOI) and the spin/orbit Hall effects (SHE/OHE) within the framework of optical Dirac theory, providing deeper physical insights into related optical phenomena. To this end, we extend optical Dirac theory \cite{Feng2022FourVectorOD} by incorporating antisymmetric components in the dielectric tensor. These components manifest either as a rotational vector in the spin-1 mode, associated with a helical optical path, or as a spin-2 tensor mode, corresponding to gyro-electric/magnetic media. 

In Section II, we derive the optical Dirac equation in a helical coordinate system with a rotational dipole-vector, leading to an effective Hamiltonian that explicitly captures both SAM-EOAM and IOAM-EOAM interactions. We also provide leading-order solutions to the optical Dirac equation, incorporating SAM/IOAM-EOAM interactions. By retaining longitudinal corrections, we further examine back-scattering phenomena related to the anti-photon state. In Section III, we employ the Heisenberg equation to derive the photon motion equation, introducing characteristic corrections to light-ray trajectories, which manifest as the spin Hall effect (SHE) and orbital Hall effect (OHE). Section IV explores the evolution of SAM and IOAM in the Heisenberg picture, elucidating their role in the broader SHE/OHE family. In Section V, we analyze the interaction between helicity-independent spin and EOAM, revealing behavior analogous to that of IOAM. Finally, Section VI summarizes our findings and presents concluding remarks. The extension to gyro-electric/magnetic media follows naturally, as shown in Appendix A.  

\section{OPTICAL DIRAC EQUATION IN HELICAL COORDINATES }

For light propagation in a gravitational field, the curved space background can be transformed to a linear optical medium, whose optical properties characterized by the effective dielectric tensor are fully specified by the spacetime geometries \cite{osti_7281940}. Typically, in a rotating gravitational field, a well-known optical phenomenon is the classical phase shift referred to as the Sagnac factor \cite{Sagnac_RMP.39.475}, currently explained by the dragging effect of the frame in general relativity \cite{Landau1980Classical}, which can be understood alternatively by the coupling between the rotation and the orbital angular momentum(OAM) of photons $\sim {\bf \Omega}\cdot{\bf L}$. In addition, another phenomenon is an extra polarization-dependent phase shift resulting from the vortical dragging vector, which leads to the optical activity of rotating gravitational fields \cite{cohen1968further,feng1989optical,carini_PhysRevD.46.5407}. Basically, it can also be equivalently described by the coupling between the macroscopic rotation ${\bf \Omega}$ and the internal spin ${\bf s}$ of the photon \cite{1993PhLA..173..347M,feng1989optical}. This coupling has a typical form of $\sim {\bf \Omega}\cdot{\bf s}$. 

\begin{figure}[htbp]
            \centering
            \includegraphics[height=0.2\textheight]{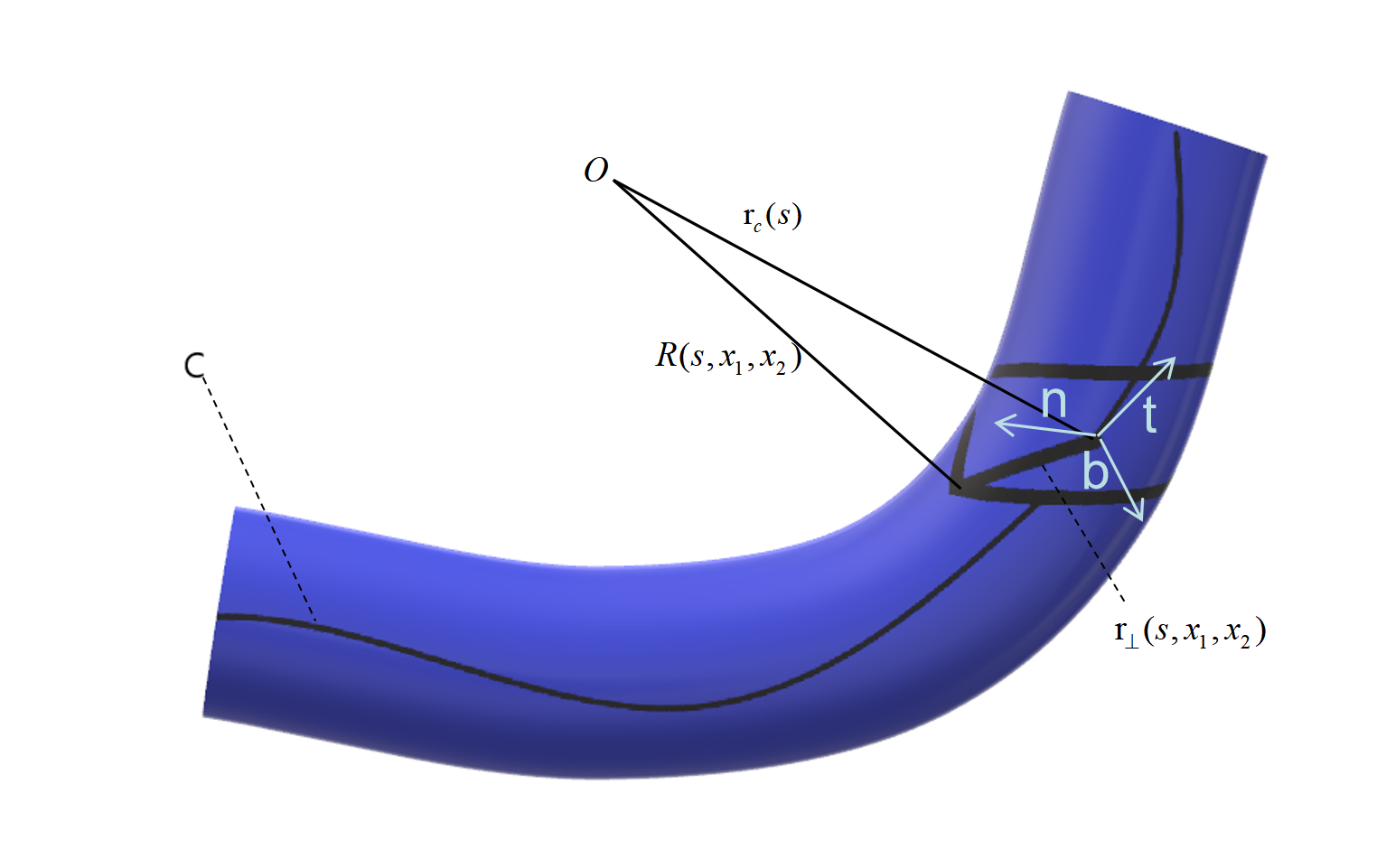}
             \caption{The local coordinate system is constructed from the Frenet frame $\mathbf{t}, \mathbf{n}, \mathbf{b}$, where $C$ denotes the trajectory of the ray system of light. The position vector $\mathbf{r}_c$ describes the ray’s location relative to the laboratory reference frame, whose origin is at point $O$. The transverse displacement vector $\mathbf{r}_{\perp}$ characterizes the deviation perpendicular to the ray direction. The blue-shaded region, spanned by the transverse displacement vectors along the ray path, defines a three-dimensional volume that can be conceptually likened to an optical fiber.}
             \label{fig1}
\end{figure}

It is noted that both rotation-OAM/SAM couplings are due to spatial rotation in time, which can be attributed to the dragging vector, the space-time component ${\bf g} = \{g_{0i}\}$ in metric. Similarly, a similar rotation occurs in space for a helical optical path. As demonstrated later, this rotation around a spatial axis will give rise to a similar coupling effect, e.g., the optical SHE. Suppose light propagates along the helical path in space (see FIG.~\ref{fig1}). Under the geometric optics approximation, an optical path can be parameterized by its optical length $s$, and let $\mathbf{r}_{\perp}=\{x_1,x_2\}$ denote a transverse displacement vector around the propagating direction; we can introduce local coordinates $(s,x_1,x_2)$, which are related to the position vector ${\mathbf R} = (x,y,z)$ in the laboratory frame by
\begin{equation}\label{localframe}
    \mathbf{R}= \mathbf{r}_c(s) + \mathbf{r}_{\perp}(s)=\mathbf{r}_c(s) + x_1{\mathbf e}_1(s)+x_2{\mathbf e}_2(s)
\end{equation}

In the local Frenet-Serret orthonormal tetrad defined by curvature $\kappa(s)$ and torsion $\tau(s)$, the line element is $d{\bf R}^2 = \widetilde{\gamma}_{\alpha\beta}dx^{\alpha} dx^{\beta}$ where Greek indices run over 0, 1 and 2, and $x_0=s$ by convention. The metric components can be easily found \cite{Takagi1992QuantumMO}:
\begin{equation}\label{eq:helicalframe}
\begin{aligned}
    \widetilde{\gamma}_{00} =(1-\kappa x_1 )^2 + \tau^2 r_{\perp}^2, \widetilde{\gamma}_{0\alpha} = -\tau \varepsilon_{0\alpha\beta} x_{\beta}, \widetilde{\gamma}_{\alpha\beta}=\delta_{\alpha\beta} 
\end{aligned}    
\end{equation}
which takes a similar form as that for a rotating frame of reference. 

In the curvilinear coordinate system, the source-free Maxwell equation can be written in the non-covariant form \cite{osti_7281940}:
\begin{equation}\label{MaxwellEq}
\begin{aligned}
&\displaystyle{\frac{1}{\sqrt{\widetilde{\gamma}}}}\frac{\partial}{\partial t}{(\sqrt{\widetilde{\gamma}}} D^{\alpha})= \epsilon^{\alpha\beta\rho}\nabla_{\beta}\bigr(\sqrt{\widetilde{\gamma}}\widetilde{\gamma}_{\rho\sigma}B^{\sigma}\bigr) \\
&\displaystyle{\frac{1}{\sqrt{\widetilde{\gamma}}}}\frac{\partial}{\partial t}{(\sqrt{\widetilde{\gamma}}} B^{\alpha})= \epsilon^{\alpha\beta\rho}\nabla_{\beta}\bigr(\sqrt{\widetilde{\gamma}}\widetilde{\gamma}_{\rho\sigma}D^{\sigma}\bigr) \\
&\displaystyle{\frac{1}{\sqrt{\widetilde{\gamma}}}}\nabla_{\alpha}(\sqrt{\widetilde{\gamma}}D^{\alpha}) =0 \\
&\displaystyle{\frac{1}{\sqrt{\widetilde{\gamma}}}}\nabla_{\alpha}(\sqrt{\widetilde{\gamma}}B^{\alpha}) =0 
\end{aligned}
\end{equation}
where $\sqrt{\widetilde{\gamma}}=|\det(\widetilde{\gamma}_{\alpha\beta})|^{1/2}$ that can be absorbed into the definition of the electromagnetic vectors ${\bf D}$ and ${\bf B}$ by this rescaling. It is noted that an electromagnetic system described by Eq.(\ref{MaxwellEq}) in curvilinear coordinates is equivalent to an optical medium with 
identical permittivity and permeability tensors, 
\begin{equation}\label{eq:delectric_helicalfibre}
\{\epsilon^{-1}\}_{\alpha\beta}=\{\mu^{-1}\}_{\alpha\beta}=\widetilde{\gamma}_{\alpha\beta}
\end{equation}
which is currently referred to as the spin-degenerate condition in photonic topological insulators. Following the same method as described in Feng \& Wu \cite{Feng2022FourVectorOD}, we project the Maxwell equation onto the helicity space spanned by ${\bf e}_{\pm}=({\bf e}_1\pm i{\bf e}_2)/\sqrt{2}$ where $\{{\bf e}_1,{\bf e}_2\}$ is given by Eq.(\ref{localframe}) to define the transverse tetrad perpendicular to the wavenumber vector of photon ${\bf e}_{\bf k}={\boldsymbol{\nabla}}_s$. Let $\bm D_{\perp}{=(D_{+},D_{-})}$, ${\bm B}_{\perp}{=(B_{+},B_{-})}$ denote the transverse components of electromagnetic field vectors in the helicity space, we introduce two-component wavefunctions for photons in the transverse helicity space:
\begin{equation}
\label{Wave Function}
{\bm\Psi}_{\pm} = {\bm D}_{\perp}\pm i\sigma_3 {\bm B}_{\perp}
\end{equation}
{\small and further combine to form a four-component wavefunction ${\bm \Psi} = ({\bm \Psi}_+, {\bm \Psi}_-)^T$}. Accordingly, the Maxwell equations in Eq.(\ref{MaxwellEq}) can be recast into a Dirac-like equation for the photon wavefunction.  

Given the permittivity and
permeability tensors Eq.(\ref{eq:delectric_helicalfibre}), following the derivation procedure outlined in Appendix \ref{app:A}, the optical Dirac equation in the helical coordinate system can be expressed as
\begin{equation}\label{eq:opticalDirac-helicalfibre}
    \begin{aligned}
i \frac{\partial \mathbf{\Psi}}{\partial t} =\Bigl[\gamma_0 \left(\hat{k}_z + \mathbf{\Omega}\cdot\mathbf{J}     \right) + \frac{1}{k_z}\bigl(\gamma_0 P_0 +\gamma_{\perp} \cdot P_{\perp}\bigr) \Bigr] \mathbf{\Psi}
    \end{aligned}
\end{equation}
in which the first term in $\gamma_0$ term of r.h.s. gives the leading order of the Hamiltonian, including the conventional momentum operator $\hat{k}_z=-i\frac{\partial}{\partial z}$ in the propagating direction and the rotation-angular momentum coupling, ${\mathbf \Omega}\cdot{\mathbf{J}}$, where the rotation is specified by the torsion of the helical path, $\mathbf{\Omega} = -\tau {\bf e}_z$, and ${\bf J}$ is the TAM, ${\bf J}={\bf L} +{\bf \Sigma} $, here ${\bf L}=\hat{\bf x}\times\hat{\bf k}$ is the OAM operator, ${\bf \Sigma} = \boldsymbol{\sigma}\otimes \sigma_0$ is the SAM operator; The second term $\propto 1/k_z$ results from the paraxial corrections, and the explicit form of $P_0$ and $P_{\perp}$ are given by Eqs.({\ref{eq:helical_momenta_A}}-\ref{eq:helical_momenta_B}) in Appendix \ref{app:A}. 

Our Hamiltonian includes non-Hermitian terms, which carry physical interpretations related to damping effects caused by interactions with the medium \cite{Graefe2009ClassicalLO}. This is particularly analogous to the behavior of evanescent waves propagating along the $z$-direction while decaying in the transverse plane. The corresponding wave function can be expressed as:
\begin{equation} \begin{aligned}\label{evanescent} \mathbf{\psi} = \psi_0 \exp \left(i k_z z - \kappa_x x - \kappa_y y \right), \end{aligned} \end{equation}
where the transverse wave vector is purely imaginary:
\begin{equation} \mathbf{k}_{\perp} {{=i{\boldsymbol \kappa}_{\perp}}} =i\kappa_x \mathbf{e}_x + i\kappa_y \mathbf{e}_y. \end{equation}
According to the Hamiltonian read from Eq.(\ref{eq:opticalDirac-helicalfibre}), a non-Hermitian scalar term $-i{\bf g}_{\perp}/k_z$ appears in $P_0$ (see Eq.(\ref{eq:helical_momenta_B})), where ${\bf g}_{\perp}=\frac{1}{2}{\boldsymbol\nabla}_{\perp}\tilde\gamma_{00}$. However, when coupled to the evanescent wave, this term transforms into a Hermitian wave-interaction ${\boldsymbol{\kappa}}_g\cdot {\boldsymbol{\kappa}}{_{\perp}}$ with ${\boldsymbol{\kappa}}_g={\bf g}_{\perp}/k_z$, and contributes an additional energy correction.

\section{TWISTED LIGHT IN THE OPTICAL DIRAC THEORY}
\label{Sec:solution}

The photon wave function, $\mathbf{\Psi}$ can be decomposed using a set of independent basis spinors. We make use of the current choice,  \begin{equation}
\begin{aligned}
    u_{\alpha,k}(z) &= u_{\alpha}(0)e^{-i\omega^{(+)} t+ik_z z}, \\
    v_{\alpha,k}(z) &= v_{\alpha}(0)e^{+i\omega^{(-)} t-ik_z z}, 
\end{aligned}    
    \quad \alpha=1,2
\end{equation}
where $k=\{\omega,{\bf k}\}$ is the wave 4-vector, $u_\alpha$ and $v_\alpha$ denote the positive and negative solutions respectively; $\alpha=1,2$ represent spin-up/-down states,    
\begin{equation}
u_\alpha(0)=\displaystyle{\binom{\chi_\alpha}{0}}, \quad v_\alpha(0)=\displaystyle{\binom{0}{\chi_\alpha}}, \quad \alpha=1,2
\end{equation}
with the 2-spinors $\chi_1=\displaystyle{\binom{1}{0}}$ and $\chi_2=\displaystyle{\binom{0}{1}}$.
Keeping the leading order of $O(k_z)$, we can easily see that the eigenvalues for the positive/negative energy states $u_{\alpha}$ and $v_{\alpha}$ are $\omega^{(\pm)}=\pm k$, respectively. In Dirac's relativistic theory of the electron, the negative energy state corresponds to its anti-particle, with opposite charge, namely, the positron. However, for the spin-1 photon, the corresponding antiparticle is itself. Obviously, it can be understood as moving toward the negative arrow of time, equivalently, a backward propagation solution.   

Now, we are considering the next-order perturbation, i.e., the zeroth-order $O(1)$ term in the power series of $k_z$ resulted from the AM-rotation coupling $\mathbf{\Omega}\cdot{\bf J} = \mathbf{\Omega}\cdot{\bf L}+\mathbf{\Omega}\cdot{\bf \Sigma}$. While including the spin-rotation coupling only, the energy eigenvalues of positive states will be altered to 
\begin{equation}
\omega^{(+)} = \omega_{\alpha} = k_z\mp \tau
\end{equation}
where $\alpha=1,2$ in the left hand corresponds to $\mp$ in the right hand. 
Similarly, for the negative energy state $v_{\alpha}$, we have
\begin{equation}
\omega^{(-)} = -\omega_{\alpha} = - (k_z \pm \tau)
\end{equation}

Given that the OAM is an odd operator under $time\-reversal$ transformation, if the vortex phase of the positive-energy photon with right/left-handed polarization is $e^{\pm il\varphi}$, accordingly, the negative-energy photon should be $e^{\mp il\varphi}$. Explicitly, we have 
\begin{equation}\label{eq:basis-rotationAM}
    \begin{aligned}
        u_{\alpha,{\bf k}}(z) &= u_{\alpha}(0)e^{-i\omega_{\alpha}^+ t+ik_z z}e^{\pm il\varphi}  \\ 
        v_{\alpha,{\bf k}}(z) &= v_{\alpha}(0)e^{-i\omega_{\alpha}^- t-ik_z z}e^{\mp il\varphi} 
    \end{aligned}
\end{equation} 
Hence, the corresponding eigen-energies are 
\begin{equation}
  \begin{aligned}
      \omega_{\alpha} &= k_z\mp (l\tau+\tau) = k_z \mp (l+1)\tau \\
      \omega_{\alpha} &= k_z\mp (l\tau-\tau) = k_z \mp (l-1)\tau
  \end{aligned}
\end{equation}
These relationships can be expressed compactly as:
\begin{equation}
    \begin{aligned}\label{dispersion}
        \omega_{\alpha} = k - j_z \tau 
    \end{aligned}
\end{equation}
with $j_z $ representing the component of the TAM $\mathbf{J} $ along the propagating direction. The possible values of $j_z$ are $j_z \in\{-l-1,-l+1,l-1,l+1\}$. 

The second term contributes to Berry phase as \cite{Bliokh2006GeometricalOO}:
\begin{equation}
    \begin{aligned}\label{Berry phase}
\gamma(C) = - j_z \gamma_B
    \end{aligned}
\end{equation}
where $\gamma_B =\int \tau ds $. This result aligns with the construction of the higher-order Poincaré sphere (HOPS) representation, as detailed in \cite{PhysRevLett.107.053601,PhysRevLett.108.190401}.

In terms of basis functions Eq.(\ref{eq:basis-rotationAM}), the 4-vector wavefunction of the photon can be written as:
\begin{equation}
    \mathbf{\Psi} = \sum_{\alpha} \bigl[a_{\alpha,{\bf k}}({\bf x}_{\perp}, z)u_{\alpha,{\bf k}}(z) + b_{\alpha,-{\bf k}}({\bf x}_{\perp},z)v_{\alpha,-{\bf k}}(z)\bigr]
\end{equation}
By substituting this wavefunction into the Dirac-like equation (\ref{eq:opticalDirac-helicalfibre}) and using the dispersion relation (\ref{dispersion}), the evolution of the photonic wavefunction can be further simplified into an optical Schrödinger-like equation \cite{Kuratsuji1997MaxwellSchrdingerEF}:
\begin{equation}
    \begin{aligned}\label{29}
       &( i \partial_z + \frac{g_{s} k^2_{\perp} }{k_z}+ \frac{1}{k_z}\hat{h}_+)  a_{1,{\bf k}}  + \frac{1}{k_z} P_+ b_{2, -{\bf k}} = 0 \\ 
       & ( i \partial_z + \frac{g_{s} k^2_{\perp} }{k_z} + \frac{1}{k_z}\hat{h}_+) b_{2, -{\bf k}} + \frac{1}{k_z} P_-a_{1,{\bf k}}  = 0
    \end{aligned}
\end{equation}
where $P_{\pm}$ is given by Eq.(\ref{eq:helical_momenta_B}), and
\begin{equation}
\hat{h}_+ = - i{\bf g}_{\perp}\cdot \mathbf{k}_{\perp} + ({\bf g}_{\perp}\times{\bf k}_{\perp})\cdot{\bf e}_{\bf k}
\end{equation}
In this formulation, the anti-photon can be interpreted as a photon propagating in the opposite direction. The transformation associated with the anti-photon can be expressed as  $b_{\alpha,\mathbf{k} } ({\bf x}_{\perp}) e^{-ikz} \rightarrow b_{ \alpha,\mathbf{-k} } ( {\bf x}_{\perp}) e^{ ikz} $.

Defining a 2-spinor through rearrangement, 
\begin{equation}
\xi = \binom{a_{1,{\bf k}}}{b_{2, -{\bf k}}}
\end{equation}
we can arrive at the following compact form, 
\begin{equation}\label{eq:2-spinor-xi}
ik_z\frac{\partial \xi}{\partial z} = - \frac{1}{2}g_{s}\nabla_{\perp}^2 \xi - \hat{h}_+\xi - \mathbf{P}_{\perp}\cdot{\boldsymbol\sigma}_{\perp}\xi
\end{equation}
in which $\mathbf{P}_{\perp}\cdot{\boldsymbol\sigma}_{\perp} = P_+\sigma_+ + P_-\sigma_-$. Similarly, defining 
\begin{equation}
\eta = \binom{b_{1,-{\bf k}}}{a_{2, {\bf k}}}
\end{equation}
results in an equation for $\eta$ with the same form as Eq.(\ref{eq:2-spinor-xi}) for $\xi$, 
\begin{equation}\label{eq:2-spinor-eta}
ik_z\frac{\partial \eta}{\partial z} = - \frac{1}{2}g_{s}\nabla_{\perp}^2 \eta - \hat{h}_-\eta - \mathbf{P}_{\perp}\cdot{\boldsymbol\sigma}_{\perp}\eta
\end{equation}
with
\begin{equation}
\hat{h}_- = -i{\bf g}_{\perp}\cdot \mathbf{k}_{\perp} - ({\bf g}_{\perp}\times{\bf k}_{\perp})\cdot{\bf e}_{\bf k}
\end{equation}

To understand the spin precession described by the interaction term ${\bf P}_{\perp}\cdot{\boldsymbol\sigma}_{\perp}$ of Eq.(\ref{eq:2-spinor-xi}) and Eq.(\ref{eq:2-spinor-eta}), it is helpful to emphasize the following three simple key points \cite{Feng2022FourVectorOD}: (1) The matrices $\sigma_{\pm}$ describe spin-flip transitions. Specifically, $\sigma_+$ induces a transition from the spin-down to the spin-up state, corresponding to the absorption of two units of angular momentum by the photon. Conversely, $\sigma_-$ induces the reverse transition, from spin-up to spin-down. (2) In the cylindrical coordinates $\{\rho,\phi,z\}$, we have 
\begin{equation}
\hat{k}_{ \pm}=\frac{1}{\sqrt{2}}(\hat{k}_x\mp i\hat{k}_y)=-\frac{i}{\sqrt{2}} e^{\mp i \phi}\left(\partial_\rho \pm \frac{1}{\rho} L_z\right) 
\end{equation}
For the eigenstate of z-component of OAM, $L_z|l\rangle=l|l\rangle$, $k_{\pm}$ acts as the ladder operators to lower/raise the z-component of orbital angular momentum (OAM) by one quantum of $\hbar$. (3) Any arbitrary dielectric tensor can be decomposed into symmetric and anti-symmetric parts. The symmetric part can further be decomposed into scalar, vector, and tensor modes, each possessing two degrees of freedom. These correspond to virtual excitation modes with spin-0 (monopole), spin-1 (dipole), and spin-2 (quadrupole), respectively. Such a spin-mode classification can be found in metric gravity theories for distinguishing distinct polarization modes of gravitational waves \cite{PhysRevLett.30.884}. While for the anti-symmetric component, it can be interpreted as inducing an effective rotational behavior in the medium, leading to the optical activity observed in gyrotropic media.

\begin{figure}[htbp]
            \centering
            \includegraphics[height=0.28\textheight]{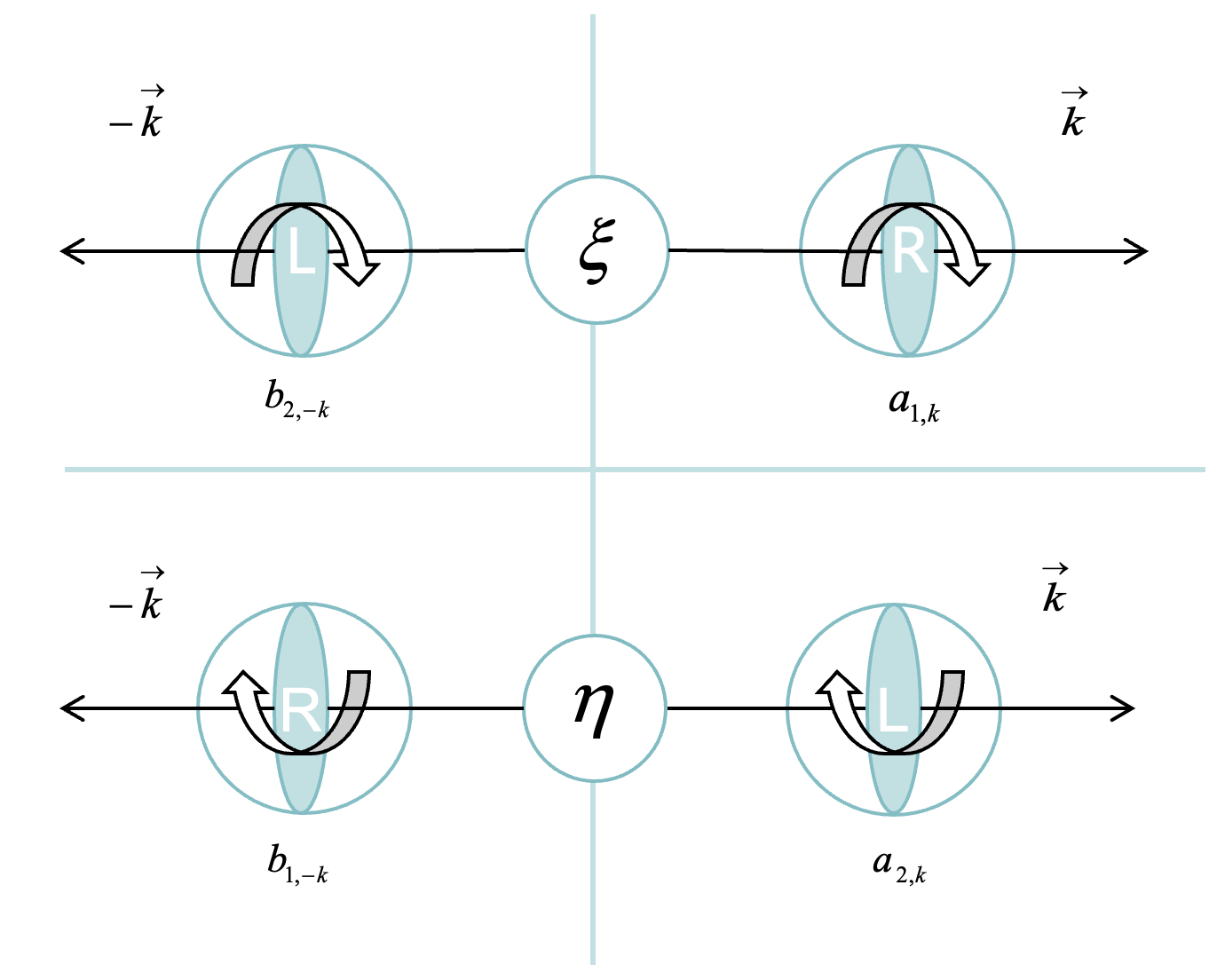}
             \caption{Illustrative diagram of photon and anti-photon helicity states. The top section illustrates two components of the 2-spinor $\xi$: $a_{1,k}$ denotes the photon traveling in the $\vec{k}$ direction with right-handed polarization, whereas $b_{2,-k}$ corresponds to its mirror state - the anti-photon - with reversed wavenumber and polarization. Similarly, the lower section illustrates two components of the 2-spinor $\eta$ but with the opposite helicity.}
             \label{fig:helicitystate}
\end{figure}

With these rules in place, the structure of the interaction term $P_{\pm} \cdot \sigma_{\pm}$ becomes more transparent, explicitly ensuring the conservation of total angular momentum (TAM) in each term and elucidating the underlying mechanism for the interconversion between different forms of angular momentum. For instance, in the first SAM-OAM interaction term proportional to $k_{\pm}^2 \sigma_{\pm}$ (see Eq.(\ref{eq:helical_momenta_B})), a spin-flip is accompanied by a change in IOAM of the photon. A flip from spin-down to spin-up (via $\sigma_+$) requires the photon to gain two units of SAM, which is compensated by a reduction of two units in OAM and vice versa for $\sigma_-$. While for the second term, it is of the dipole-IOAM-SAM interaction form $\sim d_{\pm}k_{\pm}\sigma_{\pm}$, where $d_{\pm}$ relies on the dielectric tensor can be read from Eq.(\ref{eq:helical_momenta_B}). Clearly, the compensation of two units of SAM through $\sigma_{\pm}$ is achieved by one unit of IOAM via the $k_{\pm}$ operator, with the remaining unit provided by the optical medium through the absorption of a spin-1 vector mode $d_{\pm}$. In the latter, the intervention of spin-1 modes of the medium leads to the change of its angular momentum, which provides a microscope explanation of the optical torque exerted by twisted light in the context of field theory.

In addition, we notice that there are another terms $h_{+}$ in Eq.(\ref{eq:2-spinor-xi}) for $\xi$ and $h_-$ in Eq.(\ref{eq:2-spinor-eta}) for $\eta$. It is easy to find that this spin-independent scalar term can be written in a compact form,
\begin{equation}\label{eq:h_dipole}
h_{\pm}= g_{\mp}\hat{k}_{\pm} 
\end{equation}
where $g_{\pm}$ is the components of ${\bf g}_{\perp}$ in the helicity space, i.e. ${\bf g}_{\perp}=g_\alpha{\bf e}_{\alpha}, \alpha=+,-$. Eq.(\ref{eq:h_dipole}) describes a dipole interaction between the IOAM of twisted light and optical medium. Specifically, it corresponds to the exchange of one unit of OAM quantum number, thereby imparting an additional optical torque on the medium.

Unlike the conventional two-spinor formalism used for spin-1/2 particles, the spinors $\xi$ and $\eta$ describe two-component states of photon–antiphoton pairs with fixed positive and negative helicities, respectively. Figure~\ref{fig:helicitystate} illustrates the schematic representation of the helicity states of a photon and its mirror counterpart—the antiphoton. In general, within spin-degenerate media, the helical states of photons propagate independently. However, when the antiphoton state is taken into account, a transition between the helical states of the photon and antiphoton can occur during propagation, as indicated in Eq.~(\ref{29}). This transition stems from the fact that a right-handed helical state (represented by $\xi$) for a forward-propagating photon corresponds to a left-handed state when the photon propagates in the opposite direction. Thus, the helicity state of the antiphoton can effectively be interpreted as the opposite helicity state of the photon under time or spatial inversion—physically corresponding to backscattering. Consequently, while the antiphoton may be viewed as a reflected photon with reversed helicity, the spinor fields $\xi$ and $\eta$ remain dynamically decoupled in the absence of such backscattering processes, evolving independently and preserving their respective identities within the medium.

Subject to the conservation of TAM, the emergence of an antiphoton must be accompanied by a corresponding change in OAM relative to its companion photon. This observation leads to an intriguing phenomenological consequence: in a spin-degenerate optical medium, a structured light beam with a Gaussian profile propagating in a given direction may generate antiphoton-like modes—namely, vortex beams propagating in the opposite direction while retaining the same helicity. Conversely, another possible scenario involves vortex beams, such as vector Laguerre–Gaussian modes with $l = 1$ or $l = 2$, giving rise to antiphoton modes in the form of vortex-free optical fields. These are characterized by Gaussian-like beams with a bright central intensity—effectively representing the reversal of the typical doughnut-shaped intensity profile associated with optical vortices. 

\section{EQUATION OF MOTION}

In quantum mechanics, the Heisenberg equation describes the time evolution of an operator $\hat{\mathbf{A}}$ in a Hermitian system as follows:
\begin{equation}
      \frac{d}{dt}\langle{\mathbf{ \hat{A}}}\rangle =\frac{1}{i\hbar} \langle[\hat{\mathbf{A}},\hat{\mathcal{H}}]\rangle 
\end{equation}
In our case, however, the associated Hamiltonian is non-Hermitian. To tackle this issue, the Hamiltonian can be decomposed into Hermitian and anti-Hermitian components,
\begin{equation}
    \hat{\mathcal{H}} = \hat{\mathcal{H}}_H - i\hat{\mathcal{H}}_{A}
\end{equation}
where $\hat{\mathcal{H}}_H$ and $\hat{\mathcal{H}}_{A}$ denote the Hermitian and anti-Hermitian parts respectively. The Heisenberg equation for the expectation value of an operator, $\langle \hat{A}\rangle= \frac{\langle\mathbf{\Psi}\vert \hat{A}\vert \mathbf{\Psi} \rangle}{\langle \mathbf{\Psi}\vert \mathbf{\Psi}\rangle }$, is thus modified to
\begin{equation}\begin{aligned}\label{general EOM}
     \frac{d}{dt}{ \langle\mathbf{ \hat{A}}\rangle} &=\frac{1}{i\hbar} \langle[\hat{\mathbf{A}},\hat{\mathcal{H}}_H]\rangle -\frac{1}{\hbar}\langle\{\hat{\mathbf{A}},\hat{\mathcal{H}}_{A}\}\rangle\\ 
     &  = \frac{1}{i\hbar} \langle[\hat{\mathbf{A}},\hat{\mathcal{H}}_H]\rangle -\frac{1}{\hbar}2\Gamma^2_{A,\hat{\mathcal{H}}_{A}}
     \end{aligned}
\end{equation}
where $\{,\}$  represents the anti-commutator, and $\Gamma^2_{A,\hat{\mathcal{H}}_{A}} =\frac{1}{2}\langle\{\hat{\mathbf{A}},\hat{\mathcal{H}}_{A}\}\rangle-\langle\hat
A\rangle \langle\hat{\mathcal{H}}_{A}\rangle$. For further details, refer to \cite{Graefe2009ClassicalLO}.
It is convenient to express the system through a pair of complex canonical conjugate variables, defined as \cite{Strocchi1966COMPLEXCA}: 
\begin{equation}
    \hat{z}_i= (\hat{r}_i +i\hat{k}_i)/\sqrt{2},\quad \hat{z}^*_i = (\hat{r}_i- i\hat{k}_i)/\sqrt{2}  
\end{equation}
Accordingly, the time evolution of the expectation values $\langle\hat{z}\rangle$ and $\langle\hat{z}^*\rangle$ is then given by:
\begin{equation}
\begin{aligned}
    \frac{d}{dt}\langle \hat{z}\rangle =\frac{1}{i\hbar }\langle \frac{\partial\hat{\mathcal{H}}_H }{\partial\hat{z}^*}\rangle-\frac{1}{\hbar}\langle\frac{\partial\hat{\mathcal{H}}_{A} }{\partial\hat{z}^*}\rangle\\ 
    \frac{d}{dt}\langle \hat{z}^*\rangle =-\frac{1}{i\hbar }\langle \frac{\partial\hat{\mathcal{H}}_H }{\partial\hat{z}}\rangle-\frac{1}{\hbar}\langle\frac{\partial\hat{\mathcal{H}}_{A} }{\partial\hat{z}}\rangle
    \end{aligned}
\end{equation}
In deriving these equations, we have used the following identities\cite{Louisell1973QuantumSP}: 
\begin{equation}
[\hat{z},\hat{\mathcal{H}}_{H}]=\frac{\partial\hat{\mathcal{H}}_{H}}{\partial \hat{z}^*}\quad [\hat{z}^*,\hat{\mathcal{H}}_{H}]=\frac{\partial\hat{\mathcal{H}}_{H}}{\partial \hat{z}}
\end{equation}
and
\begin{equation}
    \{\hat{z},\hat{\mathcal{H}}_{A}\}= 2\hat{\mathcal{H}}_{A} \hat{z} + \frac{\partial\hat{\mathcal{H}}_{A}}{\partial\hat{z}^*} \quad    \{\hat{z}^*,\hat{\mathcal{H}}_{A}\}= 2\hat{\mathcal{H}}_{A}\hat{z}^* + \frac{\partial\hat{\mathcal{H}}_{A}}{\partial\hat{z}^*} 
\end{equation}
In the classical limit, the Heisenberg equations of motion for the operators $\hat{z} $ and $\hat{z}^*$ are replaced by the Hamilton equation for the canonical variables $\{z,z^*\}$; Correspondingly, the quantum commutation relation $\frac{1}{i\hbar}[ , ]$ is replaced by the classical Poisson bracket of the canonical variables $\frac{1}{i} \{,\}_{z,z^*}$ \cite{Graefe2009ClassicalLO}. We replace the operator with the corresponding average 
\begin{equation}
    \langle \hat{z} \rangle \rightarrow z, \langle \hat{z}^*\rangle\rightarrow z^*
\end{equation}

The equations of motion for a classical quantity are given by \cite{Strocchi1966COMPLEXCA}:
\begin{equation}
    i \dot{z_i} = \frac{\partial \mathcal{H} }{\partial z^*_i }, \quad i\dot{z}^*_i =- \frac{\partial \mathcal{H}}{\partial z_i}
\end{equation}
Translating into the equations of motion for $r_i$ and $p_i$ yields:
\begin{eqnarray}
    \label{equation of motion}
\dot{r}_i = \frac{\partial \mathcal{H}_H}{\partial k_i} -\frac{\partial \mathcal {H}_A}{\partial r_i}\quad \dot{k_i} = - \frac{\partial\mathcal{H}_H}{\partial r_i} -\frac{\partial \mathcal{H}_A}{\partial k_i}
\end{eqnarray}
In the Hermitian part of the Hamiltonian (\ref{eq:opticalDirac-helicalfibre}), the TAM-orbit coupling term $ \mathbf{\Omega} \cdot \mathbf{J}$ consists of two distinct contributions, one arising from the photon's spin $\sigma$ and another from its topological charge $l$, both of which will lead to a unified gauge structure in the momentum space. 

Next, we proceed to derive the equations of motion while incorporating the effects of geometric phases. To account for these effects, we adopt the interaction representation, where both quantum state vectors and operators exhibit explicit time evolution. In this framework, the evolution of the state vector is governed by the Berry connection.
The system's Hamiltonian is conventionally partitioned as
\begin{equation}
  H = H_0 + H_I  
\end{equation}
where $H_0$ governs the operator dynamics while $H_I$ dictates the state vector evolution. The development of the state vector $\vert \psi(s) \rangle$ along the trajectory of light is governed by the path-ordered evolution operator:
\begin{equation}
    \vert \psi(s)\rangle = U_I(s,s_0)\vert \psi(s_0)\rangle  
\end{equation}
which satisfies the fundamental dynamical equation:
\begin{equation}\label{Interaction equation}
    i\frac{d}{ds} U_I(s,s_0) =H_I(s)U_I(s,s_0)
\end{equation}
The formal solution admits a path-ordered exponential representation:
\begin{equation}
    U_I(s) \equiv \mathcal{P}\left( e^{-i\int_{s_0}^{s} ds H_I(s)} \right)
\end{equation}
where $\mathcal{P}$ enforces path-ordering along the optical trajectory. 
In the path integral representation, the eigenvalues of the evolution operator correspond to the sum of the geometric phase and the dynamical phase \cite{PhysRevLett.61.1687}.
Within the interaction picture, the operator $U_{I}$ governs the geometric contribution to state evolution.
For infinitesimal momentum variations $\mathbf{k}\rightarrow\mathbf{k}+ \Delta\mathbf{k}$, the overlap between adjacent eigenstates $\vert \psi_k\rangle $ in momentum-space satisfies
\begin{equation}\label{Berry connection}
    \langle \psi_{\mathbf{k}+\Delta\mathbf{k}} \vert \psi_{\mathbf{k}}\rangle = 1- i \langle \psi_{\mathbf{k}} \vert  i \Delta\mathbf{ k}\cdot \partial_{\mathbf{k}} \vert \psi_{\mathbf{k}}\rangle \approx e^{-i\mathbf{A}_k(\mathbf{k}) \cdot \Delta\mathbf{ k} }
\end{equation}
where  $\mathbf{A}_k = \langle \psi_{\mathbf{k}} \vert  i\partial_{\mathbf{k}} \vert \psi_{\mathbf{k}}\rangle $ denotes the Berry connection, governing eigenstate parallel transport in parameter space.
By employing a path integral formulation, the contribution of (\ref{Berry connection}) to the geometric phase can be expressed as\cite{PhysRevLett.61.1687}:
\begin{equation} 
U_I (s) =e^{-i\oint_\Gamma A_\mathbf{k} d\mathbf{k}} 
\end{equation}
where $\Gamma$ represents a closed trajectory in momentum space. Consequently, the term $A_\mathbf{k} \cdot \dot{\mathbf{k}}$ naturally emerges as the interaction Hamiltonian $H_I$, which governs the evolution of the state vector in momentum space.
In our system, this geometric evolution is characterized by the Berry curvature tensor\cite{Bliokh2006GeometricalOO}:
\begin{equation}\label{monopole}
    \mathbf{F}(k) =\partial_\mathbf{k}\times \mathbf{A}_k= (l+\sigma) \frac{\mathbf{k}}{k^3}
\end{equation}
which exhibits a characteristic monopole structure in momentum space and 
give rise to a non-integral topological phase - the Berry phase - as given in the equation (\ref{Berry phase}). 

In a stationary ray system, the time parameter can be exchanged with the path-length parameter $s$, such that an over-dot indicates differentiation with respect to the ray parameter $s$.
Since the motion of light along the propagation direction is described by $\dot{\mathbf{r}}_c = \mathbf{e_z} $, the two-dimensional equations of motion of $\mathbf{r} =(x,y)$ and $\mathbf{k} =(k_x,k_y) $ are given as follows:
\begin{eqnarray}
        \mathbf{\dot{r}} &=& g_s\frac{\bf k}{k_z} -\mathbf{\Omega}\times \mathbf{r} - {\bf F}(k) \times \dot{\mathbf{k}} + \mathbf{\nabla}_\mathbf{p}{\Gamma} - \mathbf{\nabla}_\mathbf{r}{\Gamma_a} + \mathbf{v}_s  \label{eq:EOM-r} 
         \\ 
    \mathbf{\dot{k}} &=& \mathbf{\Omega}\times \mathbf{k} -\mathbf{\nabla}_\mathbf{r} {\Gamma}-\mathbf{\nabla}_\mathbf{k} {\Gamma_a}+\mathbf{f}_s \label{eq:EOM-k}
\end{eqnarray}

The second term in the right-hand side of Eqn.(\ref{eq:EOM-r}) describes the Lamor precession of a photon characterized by the angular velocity $-\mathbf{\Omega}$ along the helical optical path. The third term introduces a "Lorentz force" in momentum space, giving rise to the spin (or orbital) Hall effect of light  \cite{Onoda2004HallEO,Bliokh2006GeometricalOO, Bliokh_2009}. The first term in the right-hand side of the motion equation Eqn.(\ref{eq:EOM-k}) features a "Coriolis force" term, $ \mathbf{\Omega}\times \mathbf{k}$. This additional contribution comes from the OAM of the wave packet, which arises from the spatial arrangement of photons relative to their center of mass. The directional derivative of $\Gamma_h$ and $\Gamma_a$, either in real space $\mathbf{r}$ or in momentum space $\mathbf{k}$, are analyzed by decomposing the first order of $1/k_z$ in Hamiltonians $P_0 \gamma_0 + P_{\perp} \cdot \gamma_{\perp}$ into Hermitian and non-Hermitian components $P_{i}=P^h_{i} - iP^a_{i}$. Accordingly, $\Gamma_h =\langle \mathbf{\Psi}| P^h_{\perp} \cdot \gamma_{\perp}+P^h_0 \gamma_0| \mathbf{\Psi} \rangle $ represents the Hermitian part, while $\Gamma_a =\langle \mathbf{\Psi}| P^a_{\perp} \cdot \gamma_{\perp} +P^a_0\gamma_0| \mathbf{\Psi} \rangle $ corresponds to the anti-Hermitian part. These components account for non-Abelian contributions, as discussed in \cite{PhysRevA.75.053821}. At last, the terms $\mathbf{v}_s$ and $\mathbf{f}_s$ result from the contribution of the paraxial field, explicitly, given by
\begin{eqnarray}
 \mathbf{v}_s &=& \frac{1}{k_z}\mathbf{g}_{\perp}\times \mathbf{\Sigma} \\
 \mathbf{f}_s &=& -\frac{1}{k_z}\bigl({k^2_{\perp}}\mathbf{g}_{\perp}  + \tau^2 (\mathbf{\mathbf{k}\times \mathbf{\Sigma}})_{\perp}\bigr)
\end{eqnarray}
where the $\tau^2$ terms come from the second spatial derivative of $g_s$.

The expectation value of physical observables is evaluated over quantum states characterized by the wave function, $\mathbf{\Psi}$, which encompasses contributions from both positive and negative photon energy states.  Consequently, classical physical quantities derived from the wave function must also account for the contributions of antiphotons. For instance, the mean value of the momentum operator in a given state $\mathbf{\Psi}$ can be written as:
\begin{equation}\label{k-mean}
    \mathbf{k} =\frac{\langle \mathbf{\Psi}^+ \vert \hat{\bf k} \vert \mathbf{\Psi}^+ \rangle}{\langle\mathbf{\Psi}\vert\mathbf{\Psi}\rangle}-\frac{\langle \mathbf{\Psi}^- \vert \hat{\bf k} \vert \mathbf{\Psi}^- \rangle}{\langle\mathbf{\Psi}\vert\mathbf{\Psi}\rangle}  =\frac{\langle \mathbf{\Psi} \vert \gamma^0 \hat{\bf k} \vert \mathbf{\Psi} \rangle}{\langle\mathbf{\Psi}\vert\mathbf{\Psi}\rangle} 
\end{equation}
 where the photon wave functions $\mathbf{\Psi}^{\pm}=\frac{1}{2}(1\pm \gamma_0)\mathbf{\Psi}$ represent the positive and negative energy states of the photon, namely, corresponding to the photon and its mirror - antiphoton. Equation (\ref{k-mean}) becomes evident when considering the quantization of positive and negative photon states. Consequently, the expected values of operators associated with other classical physical quantities can be defined analogously.

Since our system involves non-diagonal components in the Hamiltonian, they can be diagonalized, which naturally introduces gauge fields \cite{Bliokh2004SpinGF}. Consequently, the equations of motion can be reformulated in terms of canonical coordinates  $\mathbf{R}$ and canonical momenta $\mathbf{K}$, defined as 
\begin{equation}
    \mathbf{R}=\mathbf{r}  + \mathbf{A}_{\mathbf{k}}, \mathbf{K} =\mathbf{k} + \mathbf{A}_\mathbf{r}
\end{equation}
where 
\begin{equation}
    \mathbf{R} =-i\hbar \frac{\partial }{\partial \mathbf{p}} \quad \mathbf{K} = i\hbar \frac{\partial }{\partial \mathbf{r}}
\end{equation}
This formalism aligns with the principles of noncommutative geo\-metry, yielding the following commutation relations: 
\begin{equation}\label{non-com}
    [r_i,r_j] = i\hbar \epsilon_{ijl} F^l_{\mathbf{k}} \quad 
    [k_i,k_j]=i\hbar \epsilon_{ijl} F^l_{\mathbf{r}} 
\end{equation}
where $\mathbf{F}_{\mathbf{k}}$ and $\mathbf{F}_{\mathbf{r}}$ represent field-strength tensors associated with the gauge fields.
It is evident that the Lorentz force described in Eq.(\ref{eq:EOM-r}) results from this non-commutative relation, where $\mathbf{F}_\mathbf{k} $ 
is defined in terms of Eq.(\ref{monopole}). Thus, the Hall effect terms can also be interpreted within the framework of non-commutative geometry, as discussed in \cite{Skagerstam1987LIGHTCONEGV,skagerstam,PhysRevD.35.2383}.

\section{EVOLUTION OF SPIN AND ORBIT ANGULAR MOMENTUM}
For a Hermitian system, the angular momentum evolves according to the Heisenberg's equation of motion. Conversely, anti-Hermitian components follow the evolution governed by the anti-commutation relations \cite{Graefe2009ClassicalLO}. This section is discussing the motion equation of angular momentum driven by the Hamiltonian defined in Eq.(\ref{eq:opticalDirac-helicalfibre}).

The evolution equation for the $z$-component of the OAM is given by:
\begin{equation}
\begin{aligned}
   i\hbar \frac{d}{dt} \langle \mathbf{\Psi}\vert \hat{J}_z \vert \mathbf{\Psi}\rangle &= \langle \mathbf{\Psi}\vert \hat{J}_z \hat{\mathcal{H}} -\hat{\mathcal{H}}^{\dagger}\hat{J}_z \vert \mathbf{\Psi }\rangle
    \end{aligned}
\end{equation}

Due to the contribution of the non-Hermitian term, following from Eq.(\ref{general EOM}), the evolution of the mean angular momentum is modified to \cite{Graefe2009ClassicalLO},
\begin{equation}\label{EOM,j_z}
        i\hbar \frac{d}{dt} \langle \hat{J}_z\rangle = \langle  [\hat{J}_z,\hat{\mathcal{H}}_H]\rangle - 2i\Gamma^2_{\hat{J}_z,\hat{\mathcal{H}_A}}
\end{equation}
in which $\Gamma^2_{\hat{J}_z,\hat{\mathcal{H}}_{A}} =\frac{1}{2}\langle\{\hat{J}_z,\hat{\mathcal{H}}_{A}\}\rangle-\langle\hat{J}_z\rangle \langle\hat{\mathcal{H}}_{A}\rangle $. Using the relation
\begin{equation}\label{non-commutator}
        \{ \hat{J}_z, \hat{\mathcal{H}}_A\} =[ \hat{J}_z, \hat{\mathcal{H}}_A] + 2  \hat{\mathcal{H}}_A \hat{J}_z
\end{equation}
and substituting equation (\ref{non-commutator}) into the expression for $\Gamma^2_{\hat{J}_z,\hat{\mathcal{H}}_{A}}$, we obtain the equation for the evolution of the angular momentum as follows,
\begin{equation}\label{EOM,J_Z}
        i\hbar \frac{d}{dt} \langle \hat{J}_z\rangle = \langle  [\hat{J}_z,\hat{\mathcal{H}}]\rangle+2i\bigl[\langle{\mathcal H}_A \hat{J}_z\rangle-\langle {\mathcal H}_A \rangle\langle \hat{J}_z\rangle\bigr]
\end{equation}
The second term on the right-hand side of the above equation provides the non-Hermitian correction to the conventional Heisenberg equation applicable to Hermitian systems.
According to Eq.(\ref{eq:opticalDirac-helicalfibre}), the Hermitian and anti-Hermitian components of the Hamiltonian are, respectively: 
\begin{equation}
\begin{aligned}
    &\hat{\mathcal{H}}_H = \gamma_0(\hat{k}_z +\mathbf{J}\cdot\mathbf{\Omega})+\frac{1}{k_z}(\gamma_0 P^h_0+\gamma_{\perp}\cdot P^h_{\perp}) \\ 
    &\hat{\mathcal{H}}_A = \frac{1}{k_z}(\gamma_0 P^a_0+\gamma_{\perp}\cdot P^a_{\perp})
\end{aligned}
\end{equation}

At the leading order up to $k_z^0 \sim O(1)$, the Hamiltonian is expressed as:
\begin{equation}\begin{aligned}
    &\hat{\mathcal{H}}_H = \gamma^0 (\hat{k}_z+ \mathbf{\Omega} \cdot \mathbf{J}) + i{\bf K}^{\tau}_{\perp}\cdot {\boldsymbol \gamma}_{\perp}\\
    &\hat{\mathcal{H}}_A =0
    \end{aligned}
\end{equation}
where, ${\bf K}^{\tau}_{\perp}\cdot {\boldsymbol \gamma}_{\perp} = {K}^{\tau}_{+}\gamma_+ + {K}^{\tau}_{-}\gamma_-$, with ${K}^{\tau}_{\pm} = 2\tau r_{\pm}{k}_{\pm}$.
By focusing exclusively on the 2-spinor algebra, we streamline the intermediate calculations by temporarily using $\sigma$-matrices in place of $\gamma$-matrices. Once the primary analysis is completed, the $\sigma$-matrices can be replaced back with $\gamma$-matrices to restore the final results in their original form. 

By applying the commutation relations, 
\begin{equation}\label{eq:commuator_K} 
\begin{aligned}
&[\hat{L}_z, K^{\tau}_{\pm}]\hat{\sigma}_{\pm} = \mp 2K^{\tau}_{\pm} \hat{\sigma}_{\pm}, \\ &[\hat{\sigma}_z, K^{\tau}_{\pm} \hat{\sigma}_{\pm}] = \pm 2K^{\tau}_{\pm} \hat{\sigma}_{\pm}  
\end{aligned}
\end{equation}
it is evident that the commutator of $\hat{\mathcal{H}}_H$ with both the OAM $\hat{L}_z$ and the SAM $\hat{\sigma}_z$ cancels out, $ [\hat{J}_z, \hat{\mathcal{H}}_H] = [\hat{L}_z, \hat{\mathcal{H}}_H] +[\hat{\sigma}_z, \hat{\mathcal{H}}_H]=0$, implying that, if we ignore the contribution from the second term in the right-hand side of Eq.(\ref{EOM,J_Z}), the TAM along the propagation direction, $J_z$ is conserved at the leading order. 

However, the evolution of $\bm{\hat{\sigma}}_{\perp} $ and $\hat{\mathbf{L}}_{\perp}$ exhibits Larmor precession.  Using the commutation relations 
\begin{equation}\label{eq:spin-procession-commutator}
[\sigma_3, \sigma_{\pm}] = \pm 2\sigma_{\pm}, \quad [\hat{L}_z, \hat{L}_{\pm}] = \pm \hat{L}_{\pm}
\end{equation} 
their time evolution is given by:
\begin{equation}
    \dot{\hat{\bm{\sigma}}}_{\perp} = 2\mathbf{\Omega} \times  \hat{\bm{\sigma}}_{\perp}\quad  \dot{\hat{\bf{L}}}_{\perp} = \mathbf{\Omega} \times \mathbf{\hat{L}}_{\perp}
\end{equation}
where $\mathbf{\Omega}$ only has a $z$-component, ${\bf \Omega}=-\tau {\bf e}_z$. The occurrence of the factor of 2 is because SU(2) serves as the double cover of SO(3), meaning that the SO(3) group also admits a double-valued representation in terms of $2\times2$ matrices \cite{Bhmer2012TheGO}. The commutator Eq.(\ref{eq:spin-procession-commutator}) implies that $\dot{\hat{\bm{\sigma}}} = \hat{\bm{\sigma}}\times \mathbf{G}$ where $\mathbf{G} = (2P^a_-,2P^a_+,2\tau)$. Extending this result to 4-spinors, the expectation value of the spin operator $\mathbf{\Sigma}$ obeys 
\begin{equation}
    \begin{aligned}
        \dot{\mathbf{\Sigma}} =\mathbf{\Sigma} \times \mathbf{G}
    \end{aligned}
\end{equation}
which describes the precession of spin in an effective "magnetic field" \cite{Bliokh2007, Kuratsuji1997MaxwellSchrdingerEF}.

At the first order in $1/k_z$, the Hermitian and non-Hermitian components in the diagonal Hamiltonian read $\hat{\mathcal{H}}^{(1)}_H ={k_z}^{-1}[g_s\hat{k}_{\perp}^2 + ({\bf g}_{\perp}\times\hat{\bf k}_{\perp})_z\hat{\sigma}_z]$, and $\hat{\mathcal{H}}^{(1)}_A ={\bf g}_{\perp}\cdot\hat{\bf k}_{\perp}$, respectively. The commutators between the OAM operator $\hat{L}_z $ and $\hat{\mathcal{H}}^{(1)}_H$ and $\hat{\mathcal{H}}^{(1)}_A$ are thus expressed explicitly by
\begin{equation}
\begin{aligned}\label{antiazimuthal}
    [\hat{L}_z,\hat{\mathcal{H}}^{(1)}_H] &= \frac{1}{2k_z} \left( [L_z,g_{s}] \hat{\bf k}^2_{\perp}+\left(\boldsymbol{\nabla}_{\perp} [L_z,g_{s}] \times \hat{\bf k}_{\perp}\right)_z \hat{\sigma}_z\right) \\ 
    [\hat{L}_z,\hat{\mathcal{H}}_A^{(1)}]&=\frac{1}{2k_z} \boldsymbol{\nabla}_{\perp} [L_z,g_{s}] \cdot \hat{\bf k}_{\perp}
\end{aligned}
\end{equation}
where $g_{s}=(1-\kappa x )^2 +\tau^2 r^2 $, satisfying 
\begin{equation}
    \begin{aligned}\label{L}
[\hat{L}_z,g_{s}] = - 2 i (1-\kappa x ) \kappa y 
    \end{aligned}
\end{equation}
Consequently, the TAM $J_z$ along the propagation direction is no longer conserved at the first order of $k_z^{-1}$. This loss of conservation stems from the anisotropic $\kappa$-term in Eq. (\ref{antiazimuthal}) which breaks the azimuthal symmetry \cite{10.1117/12.317704}. However, if the system retains rotational symmetry ($\kappa=0$), it can be verified that Eq. (\ref{antiazimuthal}) reduces to zero. 

We further examine the off-diagonal Hamiltonian at the $k_z^{-1}$ order, which also consists of the Hermitian component $\mathcal{P}^h_{\pm}= 2g_{\pm}k_{\pm}$, and the non-Hermitian component $\mathcal{P}^a_{\pm} = g_s k_{\pm}^2$. In the case of $\kappa=0$, we can find the same commutation relation as Eq.(\ref{eq:commuator_K}) for the Hermitian component,
\begin{eqnarray}
    \label{eq:orbit}
&[\hat{L}_z,\mathcal{P}^{h}_{\pm}]\hat{\sigma}_{\pm }  = \mp 2\mathcal{P}^{h}_{\pm}\hat{\sigma}_{\pm } \\ 
&[\hat{\sigma}_z,\mathcal{P}^h_{\pm} \hat{\sigma}_{\pm }] = \pm 2\mathcal{P}^h_{\pm} \hat{\sigma}_{\pm }
\end{eqnarray}
and also for the non-Hermitian component,  
\begin{eqnarray}
    \label{eq:spin}
 &[\hat{L}_z,\mathcal{P}^a_{\pm}]\hat{\sigma}_{\pm }  = \mp 2\mathcal{P}^a_{\pm}\hat{\sigma}_{\pm }\\ 
&[\hat{\sigma}_z,\mathcal{P}^a_{\pm}\hat{\sigma}_{\pm }] = \pm 2\mathcal{P}^a_{\pm} \hat{\sigma}_{\pm }
\end{eqnarray}
Accordingly, the spin and orbital components cancel each other out
for both the Hermitian and non-Hermitian components in the off-diagonal Hamiltonian. 

We observe that non-Hermitian terms arise only in the first-order correction of  $k_z^{-1}$, which can be attributed to the interaction between light and its propagating medium. Notably, these terms encompass both dipole and quadrupole interactions, facilitating angular momentum transfer between the SAM and OAM of photons and the spin-1/spin-2 excitation modes of the medium. If we approximate light propagation along helical paths as an adiabatic process, the quantum state of light remains in an instantaneous TAM eigenstate. Under this assumption, we have $\langle {\mathcal{H}}_A \hat{J}_z \rangle = J_z \langle {\mathcal{H}}_A \rangle$ , ensuring that the second term on the right-hand side of Eq.(\ref{EOM,J_Z}) vanishes identically.

Thus, under the condition $\kappa=0$, the Heisenberg equation Eq.(\ref{EOM,J_Z}) yields a self-consistent result, 
\begin{eqnarray}
    \frac{d}{dt} \langle \hat{J}_z \rangle = \frac{d}{dt} \langle \hat{L}_z +\hat{\Sigma}_z \rangle  = 0
\end{eqnarray}
This implies that the TAM $J_z$ is conserved and can be expressed as 
\begin{equation}
    J_z = j_z + L^{ext}_z = constant 
\end{equation}
where $L^{ext}_z$ includes contributions from the transverse shift, linked to both the SHE and the OHE \cite{Onoda2004HallEO, Bliokh2005ConservationOA, Hosten2008ObservationOT,PhysRevLett.103.100401,Dasgupta2006ExperimentalOO, Fedoseyev2001SpinindependentTS, Fedoseyev2008TransformationOT}.

Physically, the OAM discussed here consists of two components. One is the IOAM associated with the twisted wavefront characterized by the topological charge $l$, while the other is the EOAM attributed to the helical optical path, analogous to that of a classical point particle. 

Consequently, the operator $\hat{L}_z = -i\frac{\partial}{\partial \varphi}$ functions on the helical phase responsible for generating IOAM $L_{int} \propto l$, and also acts on the wave function's amplitude along the optical path, giving rise to an EOAM similar to that of a point particle, as described by
\begin{equation}
    \mathbf{L} = \mathbf{r}_c \times \mathbf{k}
\end{equation}
For the translation of the beam centroid, specifically the transverse shift $ \mathbf{r}_c \rightarrow \mathbf{r}_c +\mathbf{r}_{\perp}$, the longitudinal intrinsic vortex-dependent OAM is unchanged \cite{Bliokh2015TransverseAL}. However, the OAM of the point particle depends on the choice of the reference point, indicating that the EOAM experiences the shift given by 
\begin{equation}
    \mathbf{L}^{\prime}_{ext} = \mathbf{r}_{\perp} \times \mathbf{k}
\end{equation} 

Under the framework of general relativity, a similar conclusion can be derived from the analysis of the helical motion of spinning photons, where the spin encompasses both SAM and IOAM\cite{Corinaldesi1951SpinningTI,Plebaski1960ElectromagneticWI,Dixon1970DynamicsOE,Tod1976SpinningTP,Costa2011MathissonsHM}. In this context, the ray trajectory, $\mathbf{r}_c$, serves as a reference worldline representing the center of mass as observed in the laboratory frame. Momentum and angular momentum are defined relative to this worldline, with transverse coordinates denoted by $r_{\perp}$. When the center of mass is measured in a frame where $P^i = 0$, the corresponding position vector describes the transverse displacement associated with the Hall effect, $\mathbf{r}_{\perp} = {\bf r}_{\text{Hall}}$, as seen in the laboratory frame (see Eq. (32) in \cite{Costa2011MathissonsHM}). This displacement represents the beam's transverse shift during propagation \cite{Costa2011MathissonsHM, Puetzfeld2015EquationsOM}. The magnitude of the transverse displacement increases with the IOAM and spin, confirming that this effect arises due to the conservation of TAM \cite{Corinaldesi1951SpinningTI}.

\section{ INTRINSIC SPIN-ORBIT INTERACTIONS }

In the previous sections, we explored extrinsic spin-orbit effects, specifically the coupling between the SAM and IOAM of light and EOAM within a medium. Here, we turn our attention to intrinsic spin-orbit interactions, which manifest in evanescent waves through the intrinsic coupling of helicity-independent transverse spin with the wave vector \cite{Bliokh:2013xla, Bliokh2015TransverseAL,Aiello2015}. This intrinsic spin-orbit coupling is characterized by spin-momentum locking, resulting in effects analogous to the quantum SHE commonly observed in electronic systems \cite{Bliokh2015QuantumSH}.

For an evanescent wave with linear x-polarization, the electric field is represented by\cite{Bliokh:2013xla}:
\begin{equation}\label{eq:EvanescentField}
    \begin{aligned}
        \mathbf{E}=A_0 \left( \mathbf{e}_x-i \frac{\kappa_x}{k_z} \mathbf{e}_z \right) \exp \left(i k_z z-\kappa_x x\right)
    \end{aligned}
\end{equation}
where $A_0$ is a constant field amplitude and the complex wave vector is
\begin{equation}
    \begin{aligned}
        \mathbf{ k} = k \mathbf{e}_z + i\kappa_x \mathbf{e}_x
    \end{aligned}
\end{equation}
Here, $\kappa_x$ represents the decay constant in the transverse x-direction, and $k_z$ is the longitudinal component along the z-direction. Due to the transverse condition $\mathbf{\nabla} \cdot \mathbf{E} = \mathbf{k} \cdot \mathbf{E} =0 $, we obtain the imaginary component $E_z$ as:
\begin{equation}
    \begin{aligned}\label{E_z}
        E_z = - i \frac{\kappa_x }{k_z} E_x 
    \end{aligned}
\end{equation}
This term introduces a rotation of the electric field vector in the $(x, z)$ plane, leading to a transverse spin in the y-direction. The presence of both $E_x$ and $E_z$ components indicates that even a nominally linearly polarized wave in free space acquires an elliptical polarization in the evanescent field. 

In the context of classical electromagnetic fields, one can define the transverse spin density as \cite{10.1117/12.317704, Bliokh2015TransverseAL, S.J.1, Bliokh2012TransverseSO}:
\begin{equation}
    \mathbf{S} \varpropto {\text{Im}} (\mathbf{E}^*\times\mathbf{E})
\end{equation}
Accordingly, for the electric field given by Eq.(\ref{eq:EvanescentField}), the transverse spin density can be readily determined as, $s_y\propto\kappa_{red x}/k_z$ \cite{Bliokh2015TransverseAL}. Equivalently, the spin density $s_y$ can also be expressed as the expectation value of the spin operator in the wave function representation \cite{Berry_2009, Bliokh2012DualEH, Keller2005OnTT}.

The spin operator $\hat{S}_y $ acts as a generator of the SO(3) rotation group along the Y-axis and is explicitly represented as:
\begin{equation}
    \begin{aligned}
        \hat{S}_y = \left(\begin{array}{ccc}
             0 & 0 & i \\
             0 & 0  & 0 \\ 
             -i & 0 & 0
        \end{array}\right)
    \end{aligned}
\end{equation}
For an infinitesimal transformation, the rotation of the electric field $\mathbf{E}$ induced by the group element $e^{i\hat{S}_ys_y}$ can be represented by:
\begin{equation} 
    \begin{aligned}
        \left(\begin{array}{l}
             E_x (s)  \\
             E_y  (s) \\ 
             E_z  (s)
        \end{array}\right)  = \left(\begin{array}{ccc}
            1  & 0 & i s_y \\
             0 & 1 & 0 \\
             -i s_y & 0 & 1  
        \end{array}\right) \left(\begin{array}{l}
             E_x(0)  \\
             E_y (0) \\ 
             E_z (0 )
        \end{array}\right) 
    \end{aligned}
\end{equation}
Again, we have $s_y\propto\kappa_{ x}/k_z$ using Eq.(\ref{E_z}), .

We consider a more general case where evanescent waves propagate along the $z-$axis while decaying in both the $x$ and $y$ directions. The wave exhibits arbitrary polarization and amplitude, described by the following expression:
\begin{equation}
    \begin{aligned}
       \mathbf{E}= (E_0 \mathbf{e}_x + m E_0  \mathbf{e}_y + E_z \mathbf{e}_z)  e^{ik_z z - \mathbf{\kappa }_{\perp} \cdot \mathbf{r}_{\perp} }
    \end{aligned}
\end{equation}
where $m$ is a complex parameter that characterizes the polarization \cite{Azzam1977EllipsometryAP}. To determine the $z$ component, we apply the transverse condition $ \nabla \cdot \mathbf{E} =0$, leading to the following relationship in the helicity basis $\{{\bf e}_{\pm}\}$: 
\begin{equation}
    \begin{aligned}\label{TD}
      k_z E_z + i\kappa_+ E_- + i\kappa_- E_+  =0 
    \end{aligned}
\end{equation} 

On the other hand, the transverse spin is entirely independent of helicity and is given by \cite{Bliokh2012TransverseSO,Bliokh2013ExtraordinaryMA}:
\begin{equation}
    \mathbf{S}_{\perp} = \frac{\text{Re}\mathbf{k}\times \text{Im} \mathbf{k} }{(\text{Re}\mathbf{k})^2}
\end{equation}
In the helicity space, the transverse spin components are,  
\begin{equation}
    s_{+} = i \frac{\kappa_+}{k_z},\quad s_-=- i\frac{\kappa_-}{k_z}
\end{equation}
From the transverse condition (\ref{TD}), we find that the $z$-component of the electric field can also be expressed as
\begin{equation}
    \begin{aligned}
      E_z= \frac{-i}{k_z} (\kappa_+E_-+\kappa_-E_+) = i (\mathbf{S}_{\perp} \times \mathbf{E}_{\perp})_z 
    \end{aligned}
\end{equation}
When considering this decay term, the Hamiltonian in Eq.(\ref{eq:opticalDirac-helicalfibre}) corresponds to a shift associated with the transverse spin, given by $ k_{\pm}\rightarrow k_{\pm} \pm k_z s_{\pm} $. In this context, a shift in the operator $\hat{L}_z$ relative to the transverse spin is introduced as:
\begin{equation}\label{elliptic polarization} \mathbf{L} \cdot \mathbf{\Omega} \rightarrow \mathbf{L} \cdot \mathbf{\Omega} + (\mathbf{L}_{\perp} \times \mathbf{S}_{\perp}) \cdot \mathbf{\Omega}, \end{equation}
signifying the interplay between the transverse spin and the EOAM \cite{Aiello2009}. In the framework of geometric phase, elliptically polarized light undergoes corrections to the geometric phase. For instance, when circularly polarized light propagates through a Möbius strip, the accumulated phase is $\pi$. However, for elliptically polarized light with a nonzero $z$-component, the resulting phase deviates from $\pi$ \cite{12345}. This deviation arises from the second term in Eq.(\ref{elliptic polarization}), as the evanescent wave effectively exhibits characteristics akin to elliptic polarization.

\section{SUMMARY and CONCLUDING REMARKS}

In this paper, we revisit the photonic spin-Hall effect in inhomogeneous media within the framework of the four-vector optical Dirac theory, which provides a first-principles understanding of the spin-orbit interaction of light derived from the classical Maxwell equations in generic optical media. For the sake of simplicity,  we apply the spin-degeneracy condition, where the permittivity and permeability tensors satisfy $\epsilon = \mu$, and formulate the optical Dirac theory in curvilinear coordinates using the Frenet tetrad formalism. This approach transforms a stationary ray system with helical optical paths into an equivalent curved spacetime background. As a result, we find that an inhomogeneous medium can be transformed into a rotating frame of reference characterized by the torsion of light rays. It is noted that the spin-degeneracy condition leads to the decoupling of positive and negative helicity states of photons, without inducing spin-orbit conversion. The invariance of helicity is actually due to the duality symmetry inherent in Maxwell theory for spin-degenerate media.  

Furthermore, based on the optical Dirac equation, we construct a 2-spinor description of the photon field, in which the two components correspond to mirror states of each other, namely, opposite wavenumber and polarization but with the same helicity. Interestingly, the resulting 2-spinor optical Schrödinger equation includes a backscattering term, which pictures the conversion between the two mirror states, alternatively, photon and anti-photon states. This interpretation highlights the utility of this framework in studying light-matter interactions, such as calculating the optical torque exerted by light in various configurations.

The effective Hamiltonian derived from this formulation explicitly incorporates spin-orbit and orbit-orbit interaction terms, which are fundamental to explaining phenomena such as the geometric phase, SHE and OHE. For example, we demonstrate that the geometric phase and transverse displacement of light with IOAM are directly proportional to TAM \cite{Bliokh2006GeometricalOO}. Moreover, the effective Hamiltonian includes a non-Hermitian component that describes the transformation of light into matter. Additionally, we examine the evanescent-wave solution, characterized by transverse spin, and reveal its coupling with EOAM. Since the evanescent wave is equivalent to elliptically polarized light with a non-negligible $z$-component, this framework also accounts for the observed differences in the geometric phases of elliptically and circularly polarized light.

It is necessary to emphasize here that, to demonstrate SOI of light, we examine the case of light propagating through a uniform helical optical fiber. While this represents a limiting case, the underlying conclusions can be generalized to more complex configurations for two primary reasons - Firstly, SOI and the spin-Hall effect of light are typically treated within the framework of semi-classical theory, where the notion of a classical trajectory is utilized to describe subtle beam shifts. Basically, a ray system associated with electromagnetic waves is determined by a Hamiltonian structure induced from the dispersion relation. In our approach, this ray system is reformulated in terms of a local curvilinear coordinate system adapted to the light’s propagation path. Secondly, the eikonal approximation (i.e., geometric optics approximation) underpins most semi-classical treatments. On the other hand, the Frenet–Serret coordinate system provides a general mathematical framework for describing spatial curves in three dimensions. Under the eikonal approximation, it is reasonable to approximate the light’s trajectory locally as a piecewise-smooth curve with constant curvature and torsion, which is actually equivalent to the adiabatic transportation of a quantum system.

\begin{acknowledgments}

FLL is supported by the National Key R\&D Program of China through grant 2020YFC2201400 and the Key Program of NFSC through grants 11733010 and 11333008.

\end{acknowledgments}

\appendix

\section{OPTICAL DIRAC EQUATION IN HELICAL COORDINATES }
\label{app:A}
We start with the Maxwell equation in a generalized linear medium characterized by the tensors of both permittivity $\epsilon$ and permeability $\mu$. 
\begin{equation}\label{Maxwell}
\begin{array}{ll}
\displaystyle{\frac{\partial \mathbf{D}}{\partial t}}= \nabla \times (\mathbf{\mu}^{-1}\cdot\mathbf{B}),   &  \nabla \cdot \mathbf{D}=0 \\ \\
 \displaystyle{\frac{\partial \mathbf{B}}{\partial t}}= - \nabla \times (\mathbf{\epsilon}^{-1}\cdot\mathbf{D}), & \nabla \cdot \mathbf{B}=0
\end{array}
\end{equation}
where the local constitutive equations give the relation between the electromagnetic vectors and their corresponding auxiliary fields, ${\bf D} = \epsilon\cdot{\bf E}$ and ${\bf B} = \mu\cdot{\bf H}$. We will apply the same method (including the same notations) as \cite{Feng2022FourVectorOD} to derive the optical Dirac equation. We first notice that there is a common operation on electromagnetic vectors in the right side of the above Maxwell equation, $H\circ = \nabla\times(\Pi\circ)$, where the tensor $\Pi$ can be either $\epsilon^{-1}$ or $\mu^{-1}$, $\circ$ denotes any vector functions. Suppose a paraxial optical field propagates in the z-direction, then we transform the Maxwell equation to the helicity space spanned by $e_{\pm}=(e_x\pm ie_y)/\sqrt{2}$, which can be described in terms of a unitary transformation $\hat{U}$, explicitly  
\begin{equation}
\label{M}
   \hat{U}= \left(\begin{array}{ccc}
        1 & 1 & 0 \\
        i & -i & 0 \\
        0 & 0 & \sqrt{2}
        \end{array}\right)
\end{equation}
Under the transformation $\hat{U}$, $\nabla\times = {\bf k}\cdot{\bf s}$ where the momentum operator ${\bf k} = -i\nabla$ and $\{s_i\}_{jk}=-i\epsilon_{ijk}$, transforms to
\begin{equation}\label{eq:trans-curl}
   \hat{U}^{-1} (\hat{k}\cdot s) \hat{U} =\left(\begin{array}{cc}
        \hat{k_z} \sigma_3 & -\sigma_3 k_{\perp} \\
        -\left(\sigma_3 k_{\perp}\right)^{\dagger} & 0
        \end{array}\right)
\end{equation}
where $k_{\perp}=(k_+,k_-)^T$, $\hat{k}_{\pm} =(\hat{k}_x \mp i \hat{k}_y)/\sqrt{2}$ is the transverse momentum operator in helicity space. Similarly, let $\Pi$ denote $\epsilon^{-1}$ or $\mu^{-1}$, we have
\begin{eqnarray}\label{matrix}\label{eq:trans-matrix}
\hat{U}^{-1} \Pi \hat{U} &=& \left(\begin{array}{ccc}
Q_0 + Q_3 & Q_{+2}& Q_{+1}\\
Q_{-2} & Q_0 - Q_3 & Q_{-1} \\
Q_{-1} & Q_{+1}& q_0
\end{array}\right) \nonumber \\
&=& \left(\begin{array}{cc}
Q_0 {\bf I} + {\bf Q}\cdot{\sigma} & {\bf q}\\
 {\bf q}^{\dagger} & q_0\\
\end{array}\right)
\end{eqnarray}
where 
\begin{eqnarray}\label{Q-tensor}
&Q_0& = \frac{1}{2}(\Pi_{11}+\Pi_{22}) \\
&Q_{\pm 1}& = \frac{1}{\sqrt{2}}(\Pi_{13}\mp i\Pi_{23}) \\
&Q_{\pm 2}& = \frac{1}{2}\bigl[(\Pi_{11} - \Pi_{22}) \mp i(\Pi_{12}+ \Pi_{12})\bigr]\\
&Q_3& = \frac{1}{2}i(\Pi_{12}-\Pi_{21}) \\
&q_0& = \Pi_{33} \\
&{\bf q}& = \{Q_{+1}, Q_{-1}\}^{T} \\
&{\bf Q}&\cdot{\boldsymbol\sigma} = Q_{+2}\sigma_{+} + Q_{-2}\sigma_{-} + Q_3\sigma_3
\end{eqnarray}

In the previous work \cite{Feng2022FourVectorOD}, $\epsilon$ and $\mu$ were assumed to be the real spatio-homogeneous symmetric tensors.  In this paper, we extend this approach in two respects. First, both the monopole $q_0$ and the dipole vector ${\bf q}$ of the dielectric tensors are assumed to depend on transverse coordinates perpendicular to the propagating direction of light. As will be shown later, if the dipole vector is a rotational vector, e.g., explicitly, ${\bf q} = \Omega\times{\bf x}_{\perp}$, which can arise from a helical optical path, it will lead to a Berry potential. Moreover, we consider a gyro-electric/magnetic medium with imaginary antisymmetric components in the quadrupole ${\bf Q}$, characterised by $Q_3$ in the helicity space, as has been discussed by Bliokh et al.\cite{Bliokh2007}, it will also contribute to non-Abelian gauge potential. Both of them will produce a spin-dependent Lorentz-like force, thus giving rise to the photonic SHE.

Combining the transformation Eqs.(\ref{eq:trans-curl}) and (\ref{eq:trans-matrix}) yields
\begin{eqnarray}
\hat{H} & \rightarrow &\hat{U}^{-1} \hat{H} \hat{U} = \hat{U}^{-1} (\hat{\bf k}\cdot{\bf s}) \hat{U} \hat{U}^{-1} \Pi \hat{U} \\ 
 &=& \left(\begin{array}{cc}
\sigma_3\bigl(\hat{k}_z(Q_0{\bf I} + {\bf Q}\cdot{\boldsymbol\sigma}  - \hat{\bf k}_{\perp} {\bf q}^{\dagger}\bigr) & \sigma_3(\hat{k}_z{\bf q} -\hat{\bf k}_{\perp}q_0)\\
 - (\sigma_3 {\hat{k}_{\perp}}^{\dagger})(Q_0{\bf I} + {\bf Q}\cdot{\boldsymbol\sigma}) &  - (\sigma_3 {\hat{k}_{\perp}}^{\dagger} ){\bf q} \nonumber \\
\end{array}\right)
\end{eqnarray}
Thereby, the transverse components are given by
\begin{equation}\label{Hperp}
   \begin{aligned}
(\hat{H} {\bf V})_{\perp} = & \sigma_3\bigl[\hat{k}_z\bigl((Q_0{\bf I} + {\bf Q}\cdot{\boldsymbol\sigma}){\bf V}_{\perp}\bigr) - \hat{\bf k}_{\perp} \bigl({\bf q}^{\dagger} {\bf V}_{\perp}\bigr) \\ 
&+(\hat{k}_z{\bf q} - \hat{\bf k}_{\perp}q_0) V_z\bigr]
    \end{aligned}
\end{equation}

As we consider only the source-free Maxwell equation, the electromagnetic vectors are divergence-free vectorial fields, and the transversality condition $\nabla\cdot{\bf V}=0$ yields $V_{z}= -\hat{k}_z^{-1}\hat{\bf k}_{\perp}^{\dagger}{\bf V}_{\perp}$. Let $\hat{H}_{\perp}{\bf V}_{\perp} = (\hat{H} {\bf V})_{\perp}$, we have
\begin{equation}\label{qq}
\hat{H}_{\perp} = \sigma_3 \bigl[ H_0 + {\bf H}\cdot {\boldsymbol\sigma} \bigr]
\end{equation} 
in which 
\begin{equation}
    \begin{aligned}
        H_0 = & \hat{k}_z Q_0 + q_0\displaystyle{\frac{1}{\hat{k}_z}} \hat{k}_+ \hat{k}_{-} - {\bf q} \cdot \hat{\bf k}_{\perp}  \\ 
        &- \frac{i}{2\hat{k}_z}({\bf\nabla}_{\perp}q_0)\cdot{\bf k}_{\perp} + \frac{i}{2}{\bf\nabla}_{\perp}\cdot{\bf q}
    \end{aligned}
\end{equation}
and 
\begin{equation}
{\bf H}\cdot {\boldsymbol\sigma} = {\bf H}_{\perp}\cdot{\boldsymbol\sigma}_{\perp} + H_3\sigma_3
\end{equation}
with
\begin{equation}\label{eq:Hamilton_transverse}
\begin{array}{ll}
{\bf H}_{\perp} = \hat{k}_z {\bf Q}  + \displaystyle{\frac{1}{\hat{k}_z}} [q_0\hat{\bf k}_2 + {\bf k}_{q_0} + \hat{\bf k}_q] \\ \\
H_3 = k_zQ_3 + \displaystyle{\frac{1}{2\hat{k}_z}}(\nabla_{\perp} q_0\times {\bf k}_{\perp})_z -\displaystyle{\frac{1}{2}} ({\boldsymbol \nabla}_{\perp}\times{\bf q})_z  
\end{array}
\end{equation}
where
\begin{eqnarray} 
\hat{\bf k}_2 \cdot {\boldsymbol\sigma}_{\perp} &=& \hat{k}_{+}^2 \sigma_{+} + \hat{k}_{-}^2 \sigma_{-} \\
{\bf k}_{q_0} \cdot {\boldsymbol\sigma}_{\perp} &=& ( \hat{k}_{+} q_0) \hat{k}_{+}\sigma_{+} + ( \hat{k}_{-} q_0) \hat{k}_{-}\sigma_{-} \\
-\hat{\bf k}_q \cdot {\boldsymbol\sigma}_{\perp} &=& 2Q_{+1}\hat{k}_{+}\sigma_{+} + 2Q_{-1}\hat{k}_{-}\sigma_{-} \nonumber \\ 
&+& (\hat{k}_{+}Q_{+1})\sigma_{+} + (\hat{k}_{-}Q_{-1})\sigma_{-}  
\end{eqnarray}
Applying the above transverse projection to each curl-equation in the set of Maxwell equations Eqn.(\ref{Maxwell}), respectively, we obtain
\begin{equation}\label{TransMaxwell}
\begin{array}{ll}
i\displaystyle{\frac{\partial \mathcal{D_{\perp}}}{\partial t}}= (H_0^B - {\bf H}^B_{\perp}\cdot {\boldsymbol\sigma}_{\perp} + H_3^B\sigma_3) i{\sigma}_3 {\mathcal B}_{\perp}\\ \\
i\displaystyle{\frac{\partial i\sigma_3 \mathcal{B}_{\perp}}{\partial t}}= (H_0^D + {\bf  H}^D_{\perp}\cdot {\boldsymbol\sigma}_{\perp} +H_3^D\sigma_3) {\mathcal D}_{\perp}
\end{array}
\end{equation}
where $\mathcal{D}_{\perp}$ and $\mathcal{B}_{\perp}$ are the transverse components of electromagnetic vectors ${\mathbf D}$ and ${\mathbf B}$ in the helicity space, the superscript $D$ and $B$ denote the corresponding scalar quantities and matrices related to the operators $\nabla\times\epsilon^{-1}$ and $\nabla\times\mu^{-1}$, respectively. Introducing the two-vector wavefunctions of photons in the transverse helicity space,  
\begin{equation}
\label{WF}
{\bf \Psi}_{\pm} = {\mathcal D}_{\perp}\pm i\sigma_3 {\mathcal B}_{\perp}
\end{equation}
we further combine them to form a four-vector ${\bf \Psi}{_\perp} = ({\bf \Psi}_+, {\bf \Psi}_-)^T$. Based on these definitions, the Maxwell equation Eqn.(\ref{TransMaxwell}) can be converted into a compact form of the Dirac equation with the chiral extension, 
\begin{eqnarray}\label{opdirac}
i\frac{\partial {\bf \Psi}_{\perp}}{\partial t} = \Bigl[\gamma_0 (\hat{m}_{+} + \gamma_5\hat{m}_{-}) +{\boldsymbol\gamma} \cdot (\hat{\bf p}_{+}+\gamma_5\hat{\bf p}_{-})\Bigr] {\bf \Psi}_{\perp} \label{DMeqn2}
\end{eqnarray}
in which $\gamma$ are the Dirac representation of gamma matrices, the effective masses and momenta are  
\begin{eqnarray}
\hat{m}_{\pm} &=& \displaystyle{\frac{1}{2}}(H_0^D \pm H_0^B) \\
{\bf p}_{\pm} &=& ({\bf H}_{\perp}^{\pm},H_3^{\mp})^{t}
\end{eqnarray} 
with 
\begin{eqnarray}
{\bf H}_{\perp}^{\pm} &=& \displaystyle{\frac{1}{2}} ({\bf H}^D_{\perp} \pm {\bf H}^{B}_{\perp}) \\
H_3^{\pm} &=& \frac{1}{2}(H_3^{D} \pm H_3^{B})
\end{eqnarray}

Now, we apply the above procedure to obtain the optical Dirac equation in the helical coordinate frame specified by the spatial metric $\Gamma = \{\widetilde{\gamma}_{\alpha\beta}\}$ as given in Eq.(\ref{eq:helicalframe}). Under the unitary transformation $\hat{U}$, we have
\begin{equation}
  \begin{aligned}
    \hat{U}^{-1} \Gamma \hat{U} =\left(\begin{array}{ccc}
            1 & 0 & -i\tau r_{+} \\
            0 & 1 & i\tau r_{-} \\
            i\tau r_{-} & -i\tau r_{+} & g_{ss}
            \end{array}\right) \\ 
            =\left(\begin{array}{cc}
        I & -i \tau \sigma_3 r_{\perp} \\
        -\left(i \sigma_3\tau r_{\perp}\right)^{\dagger} & g_{s s}
        \end{array}\right)
  \end{aligned}        
\end{equation}
where $ r_{\pm}=(x\mp iy )/\sqrt{2}$. The effective Hamiltonian $({\bf k}\cdot{\bf s)\Gamma}$ undergoes the unitary transformation $\hat{H}\rightarrow \hat{U}^{-1} \hat{H} \hat{U} $, leading to the compact form:
\begin{equation}
    \begin{aligned}
        &\left(
        \begin{array}{cc}
        \hat{k_z} \sigma_3 & -\sigma_3 k_{\perp} \\
        -\left(\sigma_3 k_{\perp}\right)^{\dagger} & 0
        \end{array}\right)\left(\begin{array}{cc}
        I & -i \tau \sigma_3 r_{\perp} \\
        -\left(i \sigma_3\tau r_{\perp}\right)^{\dagger} & g_{s s}
        \end{array}\right)\\
        &=\left(
        \begin{array}{cc}
            \hat{k_z} \sigma_3+\sigma_3 k_{\perp}(i \sigma_3\tau r_{\perp})^{\dagger } & -i\tau \hat{k_z}r_{\perp} -\sigma_3 k_{\perp} g_{ss} \\
            -\left(\sigma_3 k_{\perp}\right)^{\dagger}+0 & -\sigma_3 k_{\perp}^{\dagger}-i \tau \sigma_3 r_{\perp}
            \end{array}
            \right)
    \end{aligned}
\end{equation}
Acting on the transverse components of $\mathbf{D}$, we obtain the transverse components under the Hamiltonian $\hat{H}$
\begin{equation}
\begin{aligned}
   (\hat{H}{\bf D})_{\perp} &= \left(k_z \sigma_3+ \sigma_3 k_{\perp} \cdot\left(i\tau \sigma_3 r_{\perp}\right)^{\dagger}\right) {\bf D}_{\perp}\\ 
   &+i \tau r_{\perp} \cdot k_{\perp}^{\dagger} {\bf D}_{\perp}+\frac{\sigma_3 k_{\perp} g_{ss}}{k_z} k_{\perp}^{\dagger} {\bf D}_{\perp}
   \end{aligned}
\end{equation}
where the angular momentum operator $\hat{L}_z$ is given by 
\begin{equation}
\hat{L}_z = i(r_-k_+-r_+k_-) = ir_{\perp}^{\dagger}\sigma_3k_{\perp}
\end{equation}
Projecting onto helicity space yields the effective Hamiltonian:
\begin{equation}  
    \hat{H}_{\perp}  =  \sigma_3 \Bigl[\hat{k}_z - \tau (\hat{L}_z+\sigma_3) + \frac{1}{\hat{k}_z}(P_0 +
    {\mathbf P}_{\perp} \cdot {\boldsymbol\sigma}_{\perp})\Bigr] 
\end{equation}
in which 
\begin{equation}\label{eq:helical_momenta_A}
\begin{aligned}
    &P_0 = P_I+P_3\sigma_3\\
    &{\mathbf P}_{\perp}\cdot{\boldsymbol\sigma}_{\perp} = P_{+}\sigma_+ + P_-\sigma_-
\end{aligned} 
\end{equation}
with 
\begin{equation}\label{eq:helical_momenta_B} 
\begin{aligned}
    &P_I = g_{s}k^2_{\perp}  -i{\bf g}_{\perp}\cdot {\bf k}_{\perp} \\
   &P_{\pm} = g_{s} k^2_{\pm} + 2i(\tau k_zr_{\pm}  + g_{\pm})k_{\pm} \\
    &P_3 = ({\bf g}_{\perp}\times {\bf k}_{\perp})_z 
\end{aligned}
\end{equation}
where $g_s=\widetilde{\gamma}_{00}$, $\mathbf{g}_{\perp} =\frac{1}{2}\nabla_{\perp} g_{s}$. The first term in $P_0$ is the conventional transverse-momentum correction in the paraxial approximation, the second gives the dipole interaction. Ladder operators $\sigma_{\pm}$ serve to flip spin by two angular momentum units, facilitating the first term in $P_{\pm}$ of Eq.(\ref{eq:helical_momenta_B}), which represents the SAM-IOAM conversion of photons. The second term explains how SAM and IOAM of photons interact with the spin-1 dipole mode of the optical medium. Each term ensures the conservation of angular momentum individually. The $P_3$ term is formally a spin-orbit coupling ${\bf L}_g\cdot{\boldsymbol \sigma}$ with ${\bf L}_g={\bf g}_{\perp}\times{\bf k}_{\perp}$. It is noted that the $1/k_z$ terms are from the paraxial approximation and are smaller than the zeroth leading term by a factor of $1/wk_z$ ($w$ is the waist size of paraxial optical fields).

While considering a gyrotropic medium, there exist antisymmetric components in the dielectric tensor, such as $\widetilde{\gamma}_{12} = -\widetilde{\gamma}_{21}= -i\Lambda$,
which, according to Eq.(\ref{eq:Hamilton_transverse}), leads to an additional rotation $\Lambda k_z$. Thus, the results regarding the helical optical path in this study can be extended to gyrotropic media simply by the replacement of $\Omega \rightarrow \Omega + \Lambda k_z$. 



%

\end{document}